\documentclass{article}
\usepackage{arxiv}
\usepackage{hyperref}
\usepackage[utf8]{inputenc} 
\usepackage[T1]{fontenc}    
\usepackage{booktabs}       
\usepackage{nicefrac}       
\usepackage{microtype}      
\usepackage{graphicx}
\usepackage{natbib}
\usepackage{doi}
\usepackage{algorithm}
\usepackage{algpseudocode}
\usepackage{multirow}
\usepackage{booktabs}
\usepackage{pdflscape}
\usepackage{rotating}
\usepackage{enumitem}
\usepackage{caption}
\usepackage{xurl}

\title{Intelligent Generation of Graphical Game Assets: A Conceptual Framework and Systematic Review of the State of the Art}

\date{February 14, 2023}

\author{\href{https://orcid.org/0000-0001-9828-7641}{\includegraphics[scale=0.06]{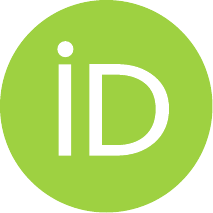}\hspace{1mm} Kaisei Fukaya}, \href{https://orcid.org/0000-0001-7849-458X}{\includegraphics[scale=0.06]{orcid.pdf}\hspace{1mm}Damon Daylamani-Zad}, \href{https://orcid.org/0000-0002-8818-2683}{\includegraphics[scale=0.06]{orcid.pdf}\hspace{1mm}Harry Agius} \\
\texttt{\{kaisei.fukaya, damon.daylamani-zad, harry.agius\}@brunel.ac.uk}\\
College of Engineering, Design and Physical Sciences\\
Brunel University London\\
Uxbridge, Middelsex, UK\\
}

\renewcommand{\shorttitle}{Intelligent graphical game asset generation: a conceptual framework and systematic review}

\begin{document}
\maketitle

\begin{abstract}
Procedural content generation (PCG) can be applied to a wide variety of tasks in games, from  narratives, levels and sounds, to trees and weapons. A large amount of game content is comprised of \textit{graphical assets}, such as clouds, buildings or vegetation, that do not require gameplay function considerations. There is also a breadth of literature examining the procedural generation of such elements for purposes outside of games. The body of research, focused on specific methods for generating specific assets, provides a narrow view of the available possibilities. Hence, it is difficult to have a clear picture of all approaches and possibilities, with no guide for interested parties to discover possible methods and approaches for their needs, and no facility to guide them through each technique or approach to map out the process of using them. Therefore, a systematic literature review has been conducted, yielding 200 accepted papers. This paper explores state-of-the-art approaches to \textit{graphical asset} generation, examining research from a wide range of applications, inside and outside of games. Informed by the literature, a conceptual framework has been derived to address the aforementioned gaps.
\end{abstract}


\section{Introduction}
Skilled artists and designers build digital content via a combination of technique and creativity. This is achieved using software such as the Autodesk products \cite{Autodesk2022}, Blender \cite{BlenderFoundation2022}, Unity engine \cite{UnityTechnologies2022} and Unreal engine \cite{EpicGames2022}. These software tools help to streamline the technical aspects of content creation, though humans still hold full creative responsibility.

Procedural content generation (PCG) is an area of research that involves the application of algorithms to the production of digital content. When specific content is desired, PCG applications can further reduce the effort and time required to make content, while retaining varying degrees of creative control. For example, SpeedTree \cite{InteractiveDataVisualizationInc.IDV2022} allow creators to specify and generate varied, high quality models of trees without the need for direct low-level manipulation. This has the effect of both optimising and democratising the content creation process. Search-based PCG methods seek out quality content by searching a content space and evaluating the results \cite{Togelius2011}, while some systems learn from existing content via machine learning (ML) \cite{summerville2018procedural}. For creative tasks, mixed-initiative approaches may be applied, allowing machine and user to co-create content \cite{Liapis2016}. These methods are popular for generating content for games, such as levels, loot \cite{BlizzardNorth2000}, puzzles or stories \cite{Bay12Games2006}. 

This work will focus on purely visual forms of content which will be referred to as \textit{graphical assets}. Graphical assets have practical usages in numerous fields, and in many cases it is necessary to use generative methods to automate all, or part of, their production. In particular, graphical assets are used in areas such as games \cite{Korn2017, Karp2021}, medicine \cite{Wang2020, Tong2020}, architectural visualisation \cite{Nishida2016, Nishida2018}, product design \cite{Alcaide-Marzal2020}, 3D printing \cite{Krs2021} and training computer-vision algorithms \cite{Jiang2018}. Though these areas use graphical assets for vastly different purposes, the methods and results need not be tied to a single application due to their visual nature. 

While existing surveys broadly examine PCG for game content \cite{Togelius2011, summerville2018procedural, Hendrikx2013}, virtual environments \cite{Smelik2014}, and deep-learning based PCG \cite{Liu2021}, none focus on the full range of purely graphical content. In addition, recent advances in deep-learning have presented impressive results in this area, with text-to-image generation \cite{Ramesh2021, Saharia2022, Rombach2022} and 3D shape learning using differentiable rendering \cite{Kato2018, Xiang2019, Henderson2020}. The body of research focuses on specific methods for generating specific assets, providing a narrow view of the available approaches and possibilities. This narrow focus makes it difficult to have a clear picture of the available approaches and possibilities. Hence, there is no guide to help interested parties to discover the possible methods and approaches for their needs, and furthermore, no facility to guide them through each technique or approach, and map out the process of using them.

This work aims to address this by aggregating state-of-the-art graphical asset generation methods. We present the results of a systematic literature review and a devised framework named \textit{Graphical Asset Generation/Transformation} (GAGeTx), for navigating these existing approaches and how to use each one.

\section{The Approach to Literature Search}
\label{sec:litReview}

The systematic literature search process is shown in figure \ref{literatureSearchProcess}. The initial search examined literature published in four main databases: ACM Digital Library, IEEE Xplore, ScienceDirect and Springer. An initial assortment of keywords was established based on general terms in the PCG literature, alongside variations and synonyms of the word “generation” or “creation”, and terms “asset” and “content”. These words, figure \ref{literatureSearchProcess}.\textit{c}, were separated into three semantic groups. 

Queries were formed by combining words from each group, the titles and abstracts of results were evaluated against the inclusion and exclusion criteria seen in figure \ref{literatureSearchProcess}.\textit{b}, and the results that passed these criteria formed the pool of accepted literature. The process of evaluation, for each query, was continued until the query was exhausted.

Queries were considered exhausted once each result had been evaluated, in the case of large searches, where results were ordered by relevance, it was necessary to deem a search exhausted once a full page of results had not passed the criteria. Some queries were too long for the ScienceDirect database, which limits the number of Boolean operators and does not accept wildcard operators. These queries had to be decomposed into smaller strings. 

The pool of accepted literature was refined by evaluating the methods and conclusions against the inclusion and exclusion criteria, then further refined by evaluating the full text against the quality criteria. At this stage, patterns emerged from the accepted literature, specifically, the common classes of graphical asset. These key graphical asset types were added as search terms, and the literature search process resumed with queries incorporating the new terms. In the accepted literature, related work was cross-referenced and evaluated against the criteria, and additional supplementary queries were performed on the databases: Ebsco, Google Scholar, and ResearchGate to ensure completeness. Pre-prints have been considered and discussed, however have not been included in frequency tables and figure \ref{literatureSearchProcess}.

\section{GAGeTx: A framework for Graphical Asset Generation/Transformation}
The accepted papers discovered through the approach presented in section \ref{sec:litReview} were examined and analysed based on the types of asset they generate, the input and output types, the techniques and approaching used. The pool of accepted papers were then grouped and categorised to derive a framework, GAGeTx, which would provide a clear picture of the current state-of-the-art and act as guide for graphical asset generation process.

\begin{table*}[ht]
\centering
    \begin{minipage}[t][][b]{.6\linewidth}
        \centering
        \includegraphics[width=\linewidth]{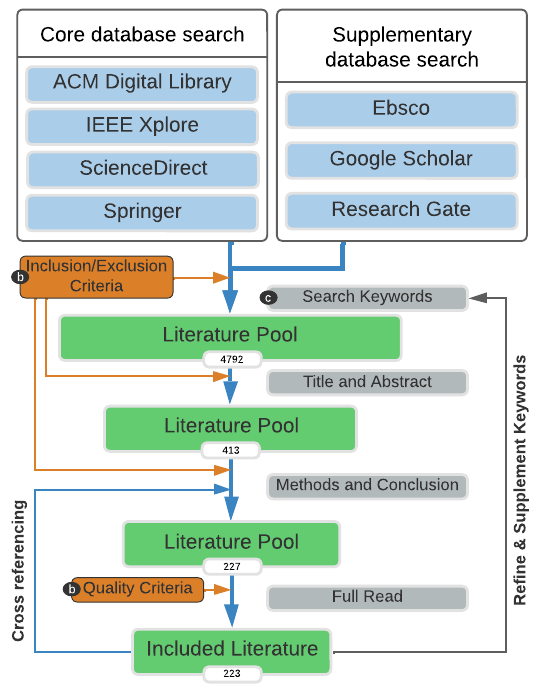}
        \label{reviewProcess}
    \end{minipage}
    \begin{minipage}[t][][t]{.35\linewidth}
      \label{incExcCriteriaTable}
      \begin{tabular}{p{\linewidth}}
        \hline
        Inclusion Criteria\\
        \hline
        \begin{itemize}[noitemsep, topsep=0pt, leftmargin=0.2cm]
            \item Methods for generating graphical assets.
            \item Studies that compare methods for generating graphical assets.
            \item Studies that combine methods for generating graphical assets.
            \item Only the newest version of a publication will be included where different iterations are found.
            \item Literature published between 2016 and 2023.
        \end{itemize}\\
        \hline
        Exclusion Criteria\\
        \hline
        \begin{itemize}[noitemsep, topsep=0pt, leftmargin=0.2cm]
            \item Studies concerning techniques for generating assets that are not distinctly graphical assets, e.g., text, audio, animation.
            \item Studies that focus on functional requirements rather than visuals.
            \item Non-procedural methods.
            \item Survey papers.
            \item Review papers.
            \item Posters.
            \item Courses.
        \end{itemize}
    \end{tabular}
    \vspace{-1cm}
    \label{qualCriteriaTable}
    \begin{tabular}{p{\linewidth}}
        \hline
        Quality Criteria\\
        \hline
        \begin{itemize}[noitemsep, topsep=0pt, leftmargin=0.2cm]
            \item The method is validated.
            \item The method is peer reviewed.
        \end{itemize}
    \end{tabular}
    \end{minipage}
    \\[-.35\baselineskip]{(a)} \hspace{0.45\linewidth} {(b)}\\
    \vspace{.5cm}
    
    \begin{minipage}[b][][b]{.45\linewidth}
      \centering
      \label{searchTermsTable1}
      \begin{tabular}{lll}
        \toprule
        \multicolumn{3}{l}{Initial search terms}\\
        \midrule
        Group 1 & Group 2 & Group 3\\
        \midrule
        Procedural* & Graphic* & Generation\\
        Algorithmic* & Asset & Synthesi* \\
        "Machine Learning" & 3D & Modeling \\
        ML & "3D Art" & Modelling \\
        "Deep Learning" & Content & Creation \\
        Inverse & "3D Model" & Design  \\
        Stochastic & Mesh & Production \\
        Grammar & Shape & Assemb* \\
        Deep &&\\
        Parametric &&\\
        \bottomrule
      \end{tabular}
    \end{minipage}
    \hspace{0.25cm}
    \begin{minipage}[b][][b]{.45\linewidth}
    \centering
      \label{searchTermsTable2}
      \begin{tabular}{lll}
        \toprule
        \multicolumn{3}{l}{Secondary set of search terms}\\
        \midrule
        Group 1 && Group 2\\
        \midrule
        Furniture & Vehicle* & “Text-to-image”\\
        Car* & Building* &\\
        Cloud* & Road* &\\
        Tree* & "Normal map" &\\
        "Texture map" & "Height map" &\\
        Environment & Terrain &\\
        Layout & Character &\\
        Face & Hair &\\
        Organ & Sprite &\\
        2D &&\\
        \bottomrule
      \end{tabular}
    \end{minipage}
    \vspace{.35cm}
    \\[-.35\baselineskip]{(c)}\\
    \captionof{figure}{Systematic literature review process: a) identifying the key databases and the return results, b) inclusion and exclusion and quality criteria applied to the literature search, c) The initial search terms used to query the chosen databases. 
    The secondary set of search terms used to query the chosen databases.} 
    \label{literatureSearchProcess}
\end{table*}

\clearpage
\begin{landscape}
    \begin{figure}
      \centering
      \includegraphics[width=\linewidth]{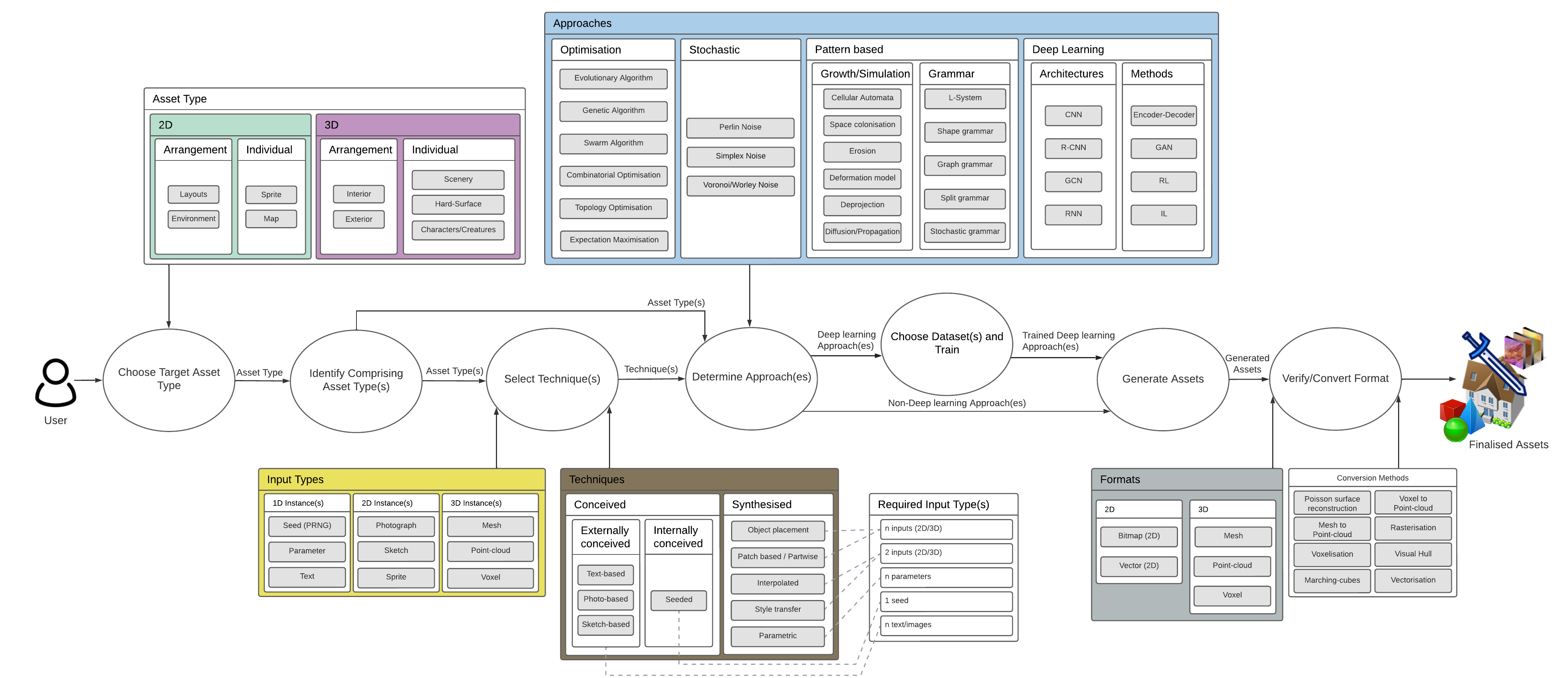}
      \caption{GAGeTx framework proposed for graphical asset generator}
      \label{frameworkFull}
    \end{figure}
\end{landscape}
\clearpage

The primary purpose of a generator is to automate or assist in a creative process. For graphical assets, a breadth of approaches may be applied to the task, depending on the desired output, level of control and available data. The GAGeTx framework, presented in figure \ref{frameworkFull}, conceptualises the task of building a generator, with respect to these aspects.

The implementation of a generator requires an understanding of the desired outcome. For instance, the user may desire variations of an existing asset, to digitise a real object via photographs, turn sketches into 3D art, or obtain quick inspiration for their own creative work. These desires can be decomposed as the type of asset required, and the technique for producing it.

Technique determines the level of control and input type of the generator. Techniques for graphical asset generation fall under two categories: \textit{conceived} and \textit{synthesised}. \textit{Conceived} techniques allow for the conception of new content either \textit{internally} from prior learning, or \textit{externally} by transforming human creative input. Conversely, \textit{synthesised} techniques construct new content by combining existing data provided at the time of generation.

The balance of creative initiative between user and generator is pertinent to the formulation of a useful generator. Conceived techniques are useful for generating new ideas or transforming a user's input to a different, useful modality, e.g. a text prompt to an image, while synthesised techniques are useful for re-configuring, or creating variations of existing content. Conceived techniques may require large datasets, and synthesised techniques may require many pieces of data from which to constitute new content, as such the choice of technique may be constricted by the availability of data. If the \textit{technique} can be seen as the task, the \textit{approach} can be seen as the solution i.e. the way in which the \textit{technique} is achieved. Different graphics formats, such as 3D meshes, point-clouds and voxels, or 2D bitmaps and vector graphics may be required for different purposes. To maximise the applicability of generative methods in cases where a different format is required, conversion methods can be applied as a final step.

The following sections will examine each step of GAGeTx in detail while reviewing the state-of-the-art. Section \ref{section-asset-types} will discuss asset types, then section \ref{section-techniques} will examine techniques. Following this, section \ref{section-approaches} will examine approaches, while discussing the bulk of the literature.  Generating and converting assets will be examined in sections \ref{section-generating} and \ref{section-converting} respectively.

\section{Asset type} \label{section-asset-types}
\begin{figure}[h]
  \centering
  \includegraphics[width=.85\linewidth]{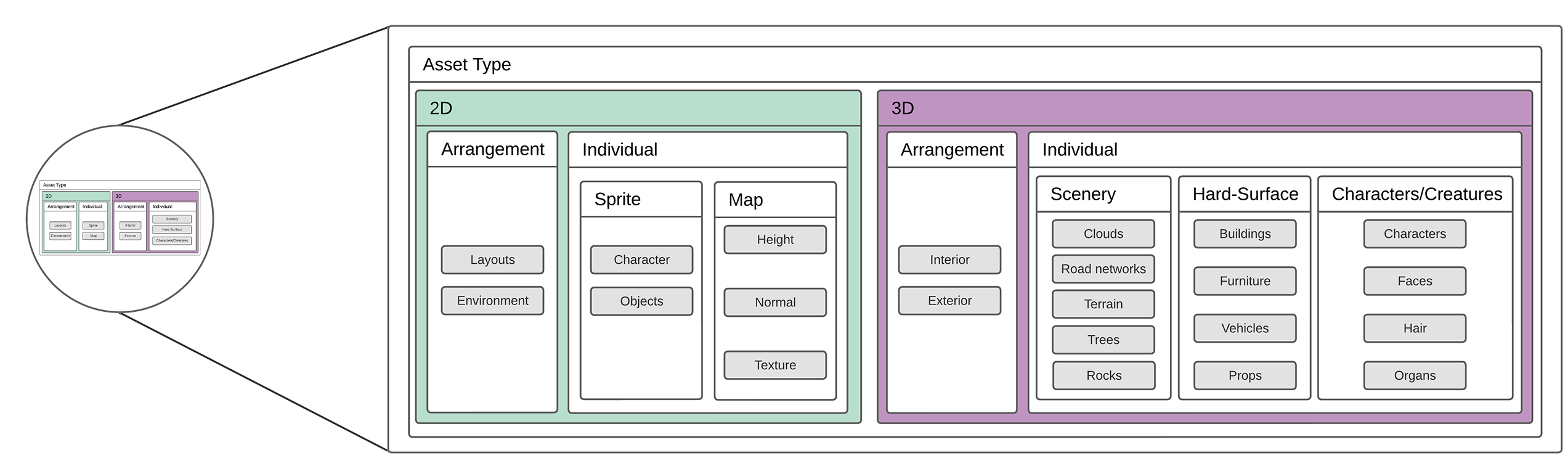}
  \caption{Asset types categorisations, populated with types observed in the literature.}
  \label{assetTypes}
\end{figure}

Each type of \textit{graphical asset} requires distinct considerations when it comes to generation. For example, structured grammar based approaches are favoured for 3D hard-surface assets, while stochastic, growth or simulation methods may be applied to scenery. This section will introduce the 21 types of asset examined in the literature and the task of selecting a target asset type within the framework. The categories of asset found in the literature are shown in figure \ref{assetTypes}, and frequency table \ref{asset-frequency} is provided to present their distribution within the literature. As evident in this table, there have been more efforts toward 3D asset generation than 2D.

\subsection{2D assets}
2D assets have applications in user-interfaces (UI) for web, print and games; presenting game worlds and characters as sprites; or augmenting 3D assets as texture, normal or height maps. With the popularity of convolutional neural networks (CNNs) in interpreting 2D data, and the development of GANs, deep learning approaches have become a large contributor in the area of 2D asset generation. In particular the Pix2Pix framework \cite{Isola2017} has had a large impact, allowing for the translation of one form of image to another. This has formed the foundation for many approaches that seek to produce content via user-sketches in particular.

\begin{table}[h]
    \centering
    \caption{Frequency and breakdown of asset types within categories. Many papers are listed here multiple times as they demonstrate the capability of producing multiple asset types.}
    \label{asset-frequency}

    \begin{tabular}[t]{c|llc}
        \toprule
        Dimension & {Asset Category} & {Asset Type} & {Frequency} \\ \toprule
        \multirow{7}{*}{2D Assets} 
        \multirow{2}{*}{} & \multirow{2}{*}{Arrangement} & Layout & 4 \\
        & & Environment & 2 \\  \cline{2-4}
        \multirow{2}{*}{} & \multirow{2}{*}{Individual (Sprite)} &     Character & 5 \\
        & & Objects & 25 \\ \cline{2-4}
        \multirow{2}{*}{} & \multirow{3}{*}{Individual (Map)} & Height & 8     \\
        & & Normal & 3 \\
        & & Texture & 8 \\ \midrule

        \multirow{14}{*}{3D Assets} 
        & \multirow{2}{*}{Arrangement} & Interior & 8 \\
        &  & Exterior & 3 \\ \cline{2-4}
        & \multirow{4}{*}{Individual (Hard-Surface)} & Buildings & 21 \\
        &  & Furniture & 51 \\
        &  & Vehicles & 39 \\
        &  & Props & 32 \\ \cline{2-4}
        & \multirow{4}{*}{Individual (Scenery)} & Cloud & 3 \\
        &  & Road network & 6 \\
        &  & Terrain & 17 \\
        &  & Tree & 6 \\ \cline{2-4}
        & \multirow{2}{*}{Individual} & Character & 17 \\
        & \multirow{2}{*}{(Characters or Creatures)} & Hair & 4 \\
        &  & Face & 12 \\
        &  & Organ & 3 \\ \bottomrule
    \end{tabular} \\

\end{table}

In print, web and UI design the arrangement of graphical elements is key to the presentation of information. Unlike sprite or 3D environments, these arrangements must be precise in order to capture the attention of an audience or user, and convey information succinctly. In web and UI design, graphical elements are not strictly visual, in many cases elements are interactive, which adds more complexity to the task of arrangement. Much like the distinction between objects and environments for 3D graphical assets, sprites can be examined individually, or as part of a larger arrangement. In 2D games many individual sprites may be arranged on the screen at once to form a cohesive game environment, these individual elements may include characters, objects and traversable areas.

Individual 2D assets take two primary forms, sprites and maps. Sprites are stand alone 2D assets, which for the purpose of this review includes general bitmap images that mimic photographs \cite{Qiao2019}, artwork \cite{Gao2020}, represent 2D characters \cite{ReboucasSerpa2019} or scenery \cite{Korn2017}.

Approaches to rendering 3D meshes make use of various 2D maps that define how a shader renders the surface of a 3D object. In modern rendering approaches, many types map may be employed including height maps \cite{Spick2019, Ivanov2020}, normal maps \cite{Su2018} and texture maps \cite{Fadaeddini2018, Gao2021}. 

\subsection{3D assets}
Among 3D assets examined in the literature, there are five main categories: interior arrangements, exterior arrangements, \textit{hard-surface}, \textit{scenery} and \textit{characters/creatures}. These are split into sub-categories, in which the \textit{hard-surface} category consists of buildings, furniture, vehicles and props, \textit{scenery} consists of clouds, road networks, terrain, trees and rocks, and \textit{characters/creatures} consists of characters, faces, hair and organs.

\textbf{3D Arrangement}: There are two distinct categories of 3D arrangement: interior and exterior. Interior arrangement refers to enclosed environments, such as bedrooms or offices; such approaches mainly emphasise object inter-relationship, where items have distinct purposes. Exterior arrangement, however, refers to the placement of objects upon a terrain, such as vegetation or buildings, where the approach to placement is more stochastic or naturalistic.

\textbf{Buildings}: The need for building generation can be found in architectural design tasks \cite{Xiong2018} and games \cite{Yoon2017,Green2019}, while in some cases entire cities are generated \cite{Waddell2002urbansim, Kim2020}. Having a simple structure, consisting of walls, doors, windows, and roofs, buildings are suited to approaches that work by combining simple elementary or parametric components, such as grammars or procedural growth-based algorithms. Some approaches specifically focus on building facades where these same techniques are applied.

\textbf{Furniture}: 3D models of furniture, such as chairs, tables, cupboards and shelves are applied commonly in architectural modelling and game worlds. Believable furniture requires a combination of functionality and stylistic consideration, specifically they must meet a functional purpose while following established design conventions. Hence, it can be beneficial to use functionality or structure aware representations when generating 3D furniture.

\textbf{Vehicles}: 3D digital environments that resemble the modern world are likely to require 3D assets that represent vehicles. The ShapeNet dataset \cite{Chang2015} contains many categories of 3D shape, including aeroplanes, buses and cars; due to the popularity of the dataset, there are many generative approaches validated on such vehicle models, including mesh \cite{Lu2020, Ling2021}, voxel \cite{Kniaz2020, Xu2021} and point-cloud \cite{Lin2021a} generation.

\textbf{Props}: To keep the list of asset types compact, hard-surface 3D objects that may be used to fill a virtual world are generically defined as props. These may include guns \cite{Wang2018}, guitars, lamps or bottles \cite{Li2021a}. As a popular dataset for generative approaches, ShapeNet \cite{Chang2015} contains many types of prop.

\textbf{Clouds}: Clouds are a common depiction in digital environments that aim to portray a realistic, earth-like world. They have a combination of attributes that make them challenging to implement, namely that they have a form but no distinct surface.

\textbf{Roads}: Roads, or road networks, are usually designed with functionality in mind. They exist to facilitate transportation throughout an environment, and in most cases can be considered in two dimensions. However, they also exist in relation to a 3D environment or terrain, conforming to a surface. To model a road system that simulates real-world roads it is also relevant to consider the human decision making involved.

\textbf{Characters}: Other than objects and terrains, digital 3D environments may be populated with varied characters, and in the case of games, characters may be customised and used as digital avatars. Alternatively, other tasks may require character mesh generation, such as checking clothing fit in online shops \cite{Abdelaziz2021}.

\textbf{Faces}: Faces are a key element of character identity, and thus, there is a need for high quality representations encompassing mesh and texture. Furthermore, in video games that allow for character customisation, the ability to adjust and customise player-character appearance is key.

\textbf{Hair}: Hair consists of many individual strands that can be of varying lengths, and flow in different ways. Current research in hair generation aims to achieve realistic hair flow, which necessitates propagation based approaches to modelling.

\textbf{Terrain}: 3D terrains have uses in various domains, from simulation, to video games and animated film. Within these domains, terrains serve the purpose of establishing a setting and environment for exploration. In games and film, terrain serves as a foundation for exterior environments, on which a digital world is built. There are two primary approaches to terrain generation, these are: surface displacement via height map, and the use of volumetric representations.

\textbf{Trees}: Many 3D digital environments make use of tree models, as shown by the popularity of SpeedTree \cite{InteractiveDataVisualizationInc.IDV2022}. Trees are the result of a natural growth process, and so tend to be visually unique. In many cases environments may require dozens if-not hundreds of trees.

\textbf{Organs}: Due to the need for accurate imaging and visualisation, the medical fields benefit from reconstructive visualisation/modelling approaches, particularly for organs. Though this does not directly relate to the other applications mentioned in this document, it is necessary to cover examples from these fields as the approaches could potentially be applied outside of that domain.

\subsection{Choose Target Asset Type and Identify their Comprising Asset Types}
The target asset type is the type of asset that the user intends to generate. It is dependant on the user's particular use case, or intended product. Within the literature many graphical asset types have been observed, though it is acknowledged that these examples do not constitute an exhaustive list of possible asset types. Techniques and approaches applied to the generation of these asset types may be applicable to similar asset types that are not already observed in the literature. As such, a generic categorisation of graphical asset types is provided in figure \ref{frameworkFull}.

A graphical asset may be comprised of multiple sub-elements, for example, a 3D furniture asset may include the model, as well as texture maps, and a character may be comprised of face, hair and body models. Producing such assets may require an approach that considers and generates all elements, or multiple approaches that handle them separately. When the target asset is an arrangement, the user may also want to generate the objects that will be arranged, in this case the objects can also be considered sub-elements, for which an appropriate generation approach must be selected.

\section{Input Types}
Within the literature, there are a variety of input types applied in the generation of assets, ranging from single value seeds, to fully formed existing 3D assets. The framework, figure \ref{frameworkFull}, presents the input types observed in the literature.

\textit{Seeds} are the simplest of input types, and are often pseudo-randomly generated. In such cases, there is no user involvement required. This simplicity, however, results in a low degree of control over the output. An example of this being basic GANs \cite{Karp2021, Fadaeddini2018, Liu2019, Wang2016}.

\textit{Parameters} provide a greater degree of control than seeds. A number of parameters can be employed, each mapping to a certain aspect of the asset, though this is dependant upon the algorithm's capacity to expose meaningful variables. Parameters can be configured by a user, but may also be pseudo-randomly generated \cite{Getto2020}, inferred using deep-learning \cite{Huang2017}, or optimised via evolutionary algorithm \cite{Muniz2021}.

Text as an input has seen recent usage in deep-learning based text-to-image transformation \cite{Ramesh2022, Rombach2022, Saharia2022}, allowing for text descriptions to be interpreted in a meaningful way to generate images. As an input, text is simple and intuitive to produce, but may not be interpreted as the user fully intends. This in some sense, asks both the user and the generator to be equally creative, or collaborative. Research into prompt engineering seeks to make this form of input more controllable \cite{Liu2022}.

Sketches are a form of input that also requires the user to be creative. Though unlike more complex input types, sketches need not be accurate or particularly detailed, requiring minimal time from a user but a good amount of creative control. Sketch based input can be interpreted either solely via deep-learning approaches \cite{Wang2020b, Delanoy2019, Kuriakose2020}, or in combination with procedural modelling \cite{Huang2017, Nishida2018}.

As inputs, point-clouds and photographs require the user to scan or photograph a subject, or otherwise source this data. The amount of control a user has over the output is constricted by the limitations of reality, that is, a subject must exist in the real-world in order for it to be scanned or photographed. Photographs are largely interpreted via deep-learning \cite{Naritomi2021, Deng2019, Lin2021b, Lei2020}.

Fully formed assets may also be used as inputs to some techniques. Such inputs, however, require the user to create precursor assets themselves or otherwise source or generate them \cite{Demir2016, Demir2017, Getto2020, Guan2020}. 

\section{Techniques} \label{section-techniques}
Technique presents the core functionality and purpose of the generator. As such, there are many possible approaches to the implementation of each technique, as will be discussed in section \ref{section-approaches}.

There are two major categories of technique. These are \emph{conceived} techniques and \emph{synthesised} techniques. Techniques are defined by the inputs they require, as well as how the data is manipulated to form a result. Conversely, the availability of inputs determines what techniques are possible. If the technique is chosen first, the input type may be derived from the chosen technique's requirements. This may not always be possible in cases where certain input data is not feasibly obtainable. In such cases a choice of input type may take precedence and the technique may be derived from this choice. 

Table \ref{technique-frequency} presents the frequency of each technique observed in the literature. As shown, the most prevalent techniques are photo-based, seeded and parametric.

\subsection{Conceived techniques}
\emph{External conception} techniques intake meaningful input from an external source, interpreting and producing a result that resembles the input. In other terms, the idea is preconceived, but the algorithm is given creative license to interpret it. Text, photo and sketch-based techniques are considered \emph{externally conceived}, as the onus is on the user to conceive of an idea, either through text prompts, photographs or hand-drawn sketches.

Text-based asset generation is explored primarily in the 2D domain, with CLIP and Diffusion based deep-learning approaches \cite{Ramesh2022, Saharia2022, Rombach2022}. Though text-based generation is achieved in 3D applications, such as with texture and displacement maps \cite{Fukatsu2021}, or full model generation \cite{Liu_2022_CVPR}. Alternatively, photo-based 3D asset generation has seen considerable exploration, with single-view \cite{Pontes2019, Naritomi2021, Lin2021b, Ren2021, Lei2020}, multi-view \cite{Lei2020}, and scan based generation of 3D assets \cite{Gao2019, Li2021}, and many utilising depth data from RGB-D images \cite{Zhang2018}. Photo-based asset generation allows for the digitisation of real-world objects, though with this comes the creative limitation that the object must exist to be photographed. In contrast, sketch-based generation allows for the creation of novel 3D \cite{Delanoy2019, Yang2021} or 2D \cite{Fukumoto2018, ReboucasSerpa2019} assets through hand-drawn designs, though different methods vary in the level of detail required from a sketch. 

\emph{Seeded} generation is considered \textit{internally conceived} as it requires no meaningful input from a user, instead producing outputs determined by a single, usually randomised value, that maps to a range of possibility. This is considered \emph{internal conception} as the algorithm conceives the output internally without meaningful input or intervention. In deep-learning, asset generation approaches commonly involve learning a latent space from a given data distribution. It is common to randomly sample from the latent space, in order to produce novel outputs \cite{Gao2021, Saito2018, tan2018variational}. In essence this random sampling is a result of noise, which is seeded using pseudo-random generation. Alternatively, some asset generation approaches may be initialised using a seed, such as in noise based algorithms for generating terrains \cite{Fischer2020, Sin2018}.

\subsection{Synthesised techniques}
In opposition to \emph{conceived} techniques, \emph{synthesised} techniques aim to produce results that are consistent with the inputs provided them. In other words the \emph{synthesised} technique performs a logical service on the given inputs, and does not provide creative input.

\emph{Object placement} involves the logical placement of pre-existing assets within a space or \emph{environment}. All \emph{arrangement} type assets, are produced via \emph{object placement}, whether for 2D layouts \cite{Li2021d, Li2021e}, 3D interiors \cite{Li2019b, Purkait2020} or 3D exteriors \cite{Zhang2019a, Sra2016}. \emph{Patch-based/Partwise} asset generation involves the piecing together of existing components into novel configurations. For instance, \cite{Teng2017} build road networks out of pre-defined patches, and \cite{Krs2021, Guan2020} piece together and morph existing meshes to form new shapes.

\emph{Interpolation} in the context of asset generation is the process of producing a result that is visually \emph{in-between} two given examples. For instance, Wang et al. \cite{Wang2018c} generate tree shapes using this kind of technique. Many generative deep-learning approaches that successfully learn a latent space, are in-turn capable of interpolation. Therefore, meaningful interpolation within the latent space is commonly used as a way of testing a GAN or VAE model for its ability to generalise \cite{Mi2021}, where successful examples show consistency in their mappings \cite{Li2021c, Gao2019a}. Hence, these approaches may implemented for the purpose of \emph{interpolation} based generation.

\emph{Style transfer} involves the application of the style of one item to the content of another. Popularised by the impactful work on neural style transfer for images \cite{Gatys2016}, many works have followed \cite{Kazemi2019, Barzilay2021, Gao2020}, including approaches to caricature \cite{Hou2021} generation and photo cartoonisation \cite{Shu2021}. Furthermore the concept of style transfer is extended into three-dimensions with mesh texturisation \cite{Hertz2020} and functionality preserving stylisation \cite{Lun2016}.

\emph{Parametric} will refer to methods that take direct numerical inputs that have meaningful affects on the output. For example, in video game character customisation a type of morphable mesh may be used \cite{Shi2022}. This may take continuous values for characteristics such as "height", "eye size" and "jaw width". By configuring these features randomly or through user input, many variations of the initial model can be generated. Alternatively, methods based on noise may take numerical inputs which directly impact the results of the noise generation \cite{Fischer2020, Satyadama2020}. Some methods aim at converting existing meshes into \emph{procedural} models \cite{Getto2020, Demir2016, Demir2017}, while others use these models as mediums for photo-based reconstruction \cite{Shi2022, Abdelaziz2021}.

\subsection{Select Techniques}
Regardless of the algorithm chosen for the task of asset generation, an input will always be required, whether this is provided by the user directly, or randomly initialised. The choice of input type primarily depends on the level of involvement and time investment the user is comfortable with.

The process of choosing the technique and input type is highly dependant on user choice. Algorithm \ref{techniqueInputAlg} presents the process for selecting a technique and input, in which the input type $input\_type_{s}$, and technique $technique_{s}$ are selected from the pool of all generative techniques $T$ and inputs $IN$.

The input type is either determined by a choice of technique, or chosen first to determine the technique. Each technique has required inputs, as seen in figure \ref{frameworkFull}. If the technique is not the priority, the input type may be chosen first, in which case the inverse limitations apply. The choice of input type requires a compromise between the user's control over the output, and the time required.

There are three methods for obtaining input data: sourcing existing data, creating data or automating the creation of data. As shown in figure \ref{inputProcess}, when data instances already exist, such as 3D meshes within ShapeNet \cite{Chang2015} or photographs found on the internet, and they are of acceptable quality and relevance, they may be used as inputs. If not, inputs can be created by the user. This provides the user with full control over the input at the cost of time and effort. Figure \ref{inputComplexity} presents the input types ordered by their complexity, with the least complex at the bottom and the most complex at the top. As the complexity of input increases the effort required in order to create, automate or source the inputs also grows.

\begin{algorithm} [h]
\caption{Selecting a Technique and Input Type}
\label{techniqueInputAlg}
	\begin{algorithmic}
	\scriptsize 
        \Procedure{Select\_Technique}{a, IN[], T[], c, u}
        \Comment{{\scriptsize INPUT: asset type, all input types, all techniques, chosen input complexity, user choices}}
            \If {$a = Arrangement$}
                \Comment{\scriptsize Arrangement assets require object placement}
                \State $technique_{s} \gets Object\_placement$
                \Comment{{\scriptsize s stands for selected}}
                \State $IN \gets [\forall \; inp \in IN \mid inp.type = dimensions ]$
                \Comment{\scriptsize Filter input type choices by asset type "dimension"}
            \Else
            \State i=0
                \While{$i < |T| \land T[i] \neq u.technique$}
                    \Comment{\scriptsize Allow user to choose an available option}
                        \State $technique_{s} \gets$ T[i]
                        \Comment{{\scriptsize technique chosen by user}}
                        \State $required\_input\_types = [\forall \; inp\in IN\mid  inp \in technique_{s}.input\_types ]$
                        \State $IN \gets$ required\_input\_types
                        \Comment{\scriptsize Filter choices by input types required by technique}
                        \State i = i +1
                \EndWhile
            \EndIf
            
            \If {$|IN|$ = 1}
                \Comment{Only one choice is available}
                \State $input\_type_{s} \gets IN[0]$
            \Else
            \State i=0
                \While{$i < |IN| \land IN[i] \neq u.input\_type$}
                    \Comment{\scriptsize Allow user to choose an available option}
                        \State $input\_type_{s} \gets$ $IN[i]$
                \EndWhile
            \EndIf
            \State return $technique_{s}, input\_type_{s}$ 
            \Comment{\scriptsize Pass selected technique and its input type to next step}
	    \EndProcedure
    \end{algorithmic}
\end{algorithm}
 
\begin{table*}
    \centering
    \begin{minipage}{.48\linewidth}
    \centering
      \includegraphics[width=\linewidth]{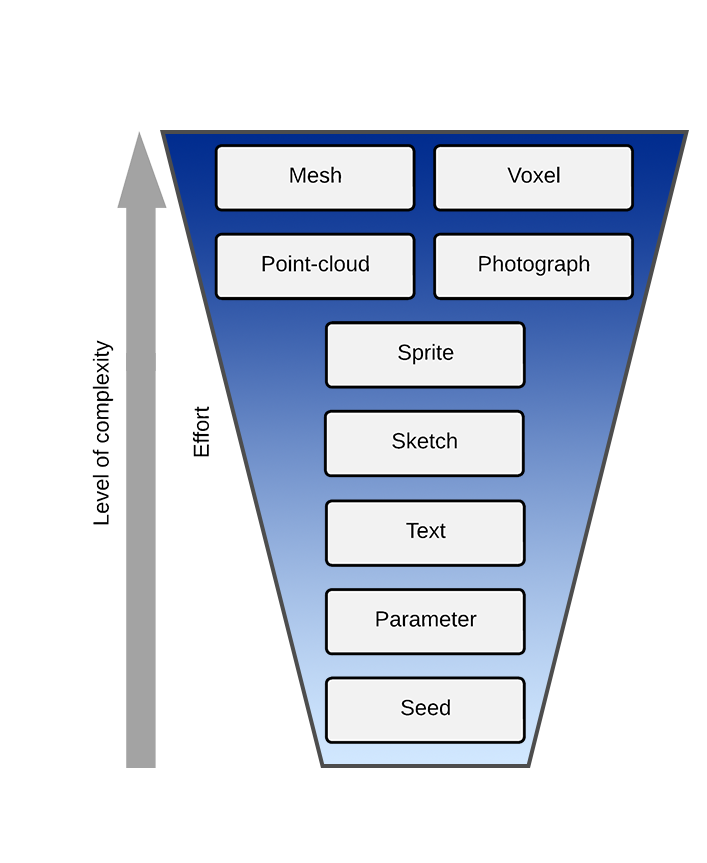}
      \captionof{figure}{Input complexity and effort diagram, going from the lowest complexity input type "seed" to the highest including "mesh" and "voxel".}
      \label{inputComplexity}
  \end{minipage}
  \begin{minipage}{.48\linewidth}
      \centering
      \includegraphics[width=\linewidth]{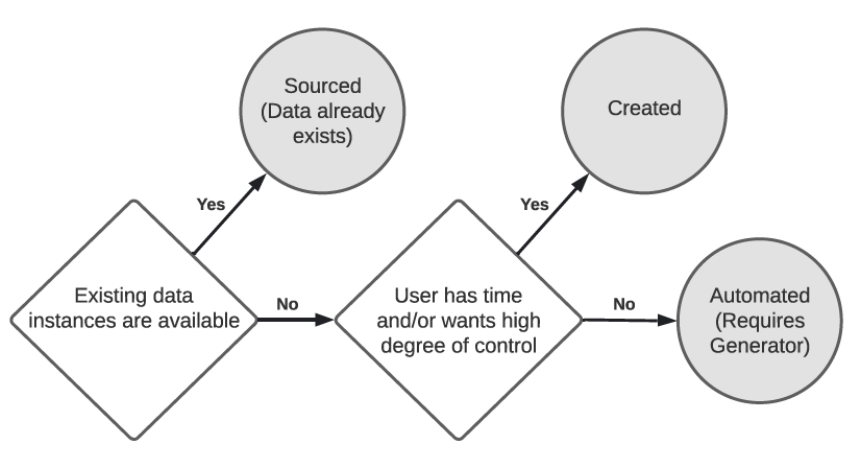}
      \captionof{figure}{Decision process for obtaining inputs.}
      \label{inputProcess}
      \vspace{.5cm}
    \centering
    \captionof{table}{Frequency and breakdown of techniques within categories. Many papers are listed here multiple times as they demonstrate multiple techniques.}
    \label{technique-frequency}
    \begin{tabular}[t]{llc}
        \toprule
        {Technique category} & {Technique} & {Frequency} \\ \toprule
        \multirow{3}{*}{Externally Conceived} 
        & Text-based & 14 \\
        & Photo-based & 66 \\
        & Sketch-based & 24 \\
        \multirow{1}{*}{Internally Conceived}
        & Seeded & 43 \\
        \midrule
        
        \multirow{5}{*}{Synthesised} 
        & Object placement & 16 \\
        & Patch based / Partwise & 4 \\
        & Interpolated & 13 \\
        & Style transfer & 12 \\ 
        & Parametric & 29 \\ \bottomrule
    \end{tabular} \\
    \end{minipage}
\end{table*}

\section{Approaches} \label{section-approaches}
After selecting a target \emph{asset type}, \emph{technique} and \emph{input type}, an appropriate \emph{approach} is determined. The \emph{approach} is the specific set of algorithms or processes that perform a specific technique, generating a target \emph{asset type} using particular \emph{inputs}. For improved presentation, approaches have been grouped and taxonomised under four main headings: \emph{optimisation}, \emph{stochastic}, \emph{pattern based} and \emph{deep learning}. This is presented in the framework, figure \ref{frameworkFull}.

Table \ref{approach-frequency} presents the frequency of each approach within the literature. Though in many cases a combination of approaches are employed, the prevalence of grammars and deep-learning for the task of generating graphical assets is made evident. In particular, many instances of shape grammars, encoder-decoder networks and GANs are observed.

\begin{table}[h]
    \centering
    \caption{Frequency and breakdown of approaches within categories. Many papers are listed here multiple times as they demonstrate multiple techniques.}
    \label{approach-frequency}

    \begin{tabular}[t]{c|llc}
        \toprule
        {Approach category} & {Approach sub-category} & {Approach} & {Frequency} \\ \toprule
        \multirow{6}{*}{Optimisation} & 
        & Evolutionary Algorithm & 2 \\
        & & Genetic Algorithm & 6 \\
        & & Swarm Algorithm & 1 \\
        & & Combinatorial Optimisation & 1 \\
        & & Topology Optimisation & 1 \\
        & & Expectation Maximisation & 1 \\ \cline{1-4}
        \multirow{3}{*}{Stochastic} &
        & Perlin Noise & 4 \\
        & & Simplex Noise & 1 \\
        & & Voronoi/Worley Noise & 1 \\ \cline{1-4}
        
        \multirow{11}{*}{Pattern based} & \multirow{6}{*}{Growth/Simulation}
        & Cellular Automata & 2 \\
        & & Space colonisation & 4 \\
        & & Erosion & 2 \\
        & & Deformation model & 3 \\
        & & Deprojection & 8 \\
        & & Diffusion/Propagation & 7 \\ \cline{2-4}
        & \multirow{5}{*}{Grammar}
        & L-System & 5 \\
        & & Shape grammar & 13 \\
        & & Graph grammar & 1 \\
        & & Split grammar & 1 \\
        & & Stochastic grammar & 2 \\ \cline{1-4}

        \multirow{8}{*}{Deep Learning} & \multirow{4}{*}{Architectures}
        & CNN & 37 \\
        & & R-CNN & 4 \\
        & & GCN & 8 \\
        & & RNN & 1 \\ \cline{2-4}
        & \multirow{4}{*}{Methods}
        & Encoder-Decoder & 30 \\
        & & GAN & 69 \\
        & & RL & 2 \\
        & & IL & 1 \\
        \bottomrule
    \end{tabular} \\

\end{table}

\subsection{Optimisation}
\textit{Optimisation} based generative approaches include \textit{evolutionary}, \textit{genetic} and \textit{swarm} algorithms, as well as \textit{combinatorial} and \textit{topology} optimisation.

\textit{Evolutionary algorithms} iteratively refine generated examples in accordance to a fitness function. At each generation, candidates with the highest fitness score are combined (through \textit{crossover}) or altered (\textit{mutated}), to produce a new generation of candidates. Over multiple generations, the fitness of candidates will improve, resulting in stronger, (high fitness) candidates. The Procedural Iterative Constrained Optimiser (PICO) framework \cite{Krs2021} is centred around a graph that represents a flow of parameterised operations that generate a 3D shape. An evolutionary algorithm is used to generate and optimise this graph, incorporating user-constraints. This is used to generate a variety of 3D assets, including trees, chairs and terrains. Functionality-Aware Model Evolution (FAME) \cite{Guan2020} evolves novel shapes in a functionality-aware manner. An evolutionary algorithm is applied to a set of models by performing crossover between groupings of parts. Users can set functionality constraints on this system, or guide evolution by selecting preferred results.
\textit{Genetic algorithms (GA)} are a popular class of evolutionary algorithm, employed in the generation of buildings \cite{Yoon2017, Merras2018}, vehicles \cite{Ban2020}, props \cite{KiptiahBintiAriffin2017, Merras2018} and clouds \cite{Muniz2021}. A CNN is used to learn a fitness function for the optimisation of cloud shapes, scoring generated clouds based on how real they appear \cite{Muniz2021}. Alternatively, \textit{GA} is applied in object placement \cite{Sra2016}. In this method, an optimal scene layout is generated based on fitness to a set of positional rules, defined by the authors. GA have also been used for camera parameter estimation \cite{Merras2018}.

\textit{Interactive genetic algorithms (IGA)} integrate user input as part of the fitness calculation, this allows a user to influence the development of the asset. The 3DCSS framework \cite{Ban2020}, for example, successfully integrates \textit{IGA} into the car design process. Alternatively, \cite{Yoon2017} attempts 3D building generation with textures using \textit{IGA}. Though, in a user study, their \textit{IGA} approach to texture generation was rated poorly, the generation of building models was effective. In another method \cite{KiptiahBintiAriffin2017}, \textit{IGA} is combined with L-systems to produce abstract 3D shapes.

\textit{Swarm algorithms} employ the use of many agents that behave independently while influenced by the group. For example, in particle swarm optimisation (PSO), a global optimum can be found in a search-space via a combination of individual search and group knowledge \cite{Kennedy1995}. 3D asteroid meshes have been generated based on real data for the purpose of simulating traversal of such terrains \cite{Li2018b}. Here, PSO is used to set optimal parameters, ensuring that the shape and surface texture of the asteroids are realistic.

\textit{Combinatorial optimisation} seeks to find optimal solutions to problems in vast but finite search spaces. The work of \cite{Lun2016} uses \textit{tabu search} to perform shape style transfer that preserves object functionality. In this method, tabu search is applied in the combinatorial optimisation of the shape such that functionality is preserved and style adaptation is maximised. This is achieved by efficiently searching through the possible modifications that can be made to the shape.

\textit{Topology optimisation} aims at producing an optimal shape within a design space based on physical constraints. This is employed in \cite{Kazi2017}, wherein 3D solutions are generated based on user provided sketches and constraints using the level-set method of \cite{ALLAIRE20021125}. This sketch based framework is effective at helping a designer to explore solutions to their specifications, though slow computation times make it less feasible for fast design iteration. A form of \textit{expectation maximisation}, first introduced by \cite{Kwatra2005} is applied in 3D cloud generation using photographs \cite{Iwasaki2017}.

\subsection{Stochastic}

\textit{Stochastic} approaches primarily involve the manipulation of noise in forming randomised yet controlled shapes and designs. Though noise underpins a large proportion of generative approaches, including many deep-learning architectures, this section will discuss methods that focus on its usage. There are many noise algorithms in common use, each with their own characteristics, these include: \textit{Perlin noise}\cite{Perlin1985}, \textit{Simplex noise}\cite{perlin_2001} and \textit{Voronoi/Worley noise}\cite{Worley1996}. Usage of noise can be \textit{seeded} and \textit{parametric} depending on the implementation or number of variables exposed to the user. Some approaches based on fractional Brownian Motion \cite{Mandelbrot1968}, for example, have parameters such as \textit{octaves}, \textit{lacunarity} and \textit{gain} that the user may adjust.

A common use case for noise is in the generation of terrains via height maps. Height maps provide a two-dimensional representation of land height which can be applied via mesh surface displacement. For example \cite{Fischer2020} employs \textit{Simplex noise} in height map terrain generation as part of a multi-step pipeline for 3D environment generation, while \cite{Li2018b} uses \textit{Simplex noise} to create surface variation on asteroid models and \cite{Satyadama2020} use \textit{Simplex noise} for surface variation when generating volumetric caves. Alternatively, \cite{Sin2018} use 3D \textit{Perlin noise} to generate consistent height mapping around a spherical surface, and \cite{Dey2018} initialise a volumetric terrain using \textit{Perlin noise}.

Many methods combine noise with other approaches as a means to reflect the roughness and variation found in nature. As an alternative to height map generation, \cite{Becher2017} introduce a method and pipeline for generating terrains using feature curves in volumetric space. The application of the feature curves concept to 3D volumetric space is an extension of previous applications to 2D height maps \cite{hnaidi:hal-01381590}. Extending to volumetric space allows for overhangs and tunnels in terrain. Montenegro et al. \cite{Montenegro2017} propose a method, based on \cite{Lipus2005}, which combines the use of implicit modelling and noise in generating clouds. This implementation performs in real-time, allowing for rapid iteration on ideas. 2D content can also be generated using noise. For example, texture maps for colouring the walls of caves have been generated using \textit{Perlin} and \textit{Worley} noise \cite{Franke2021}, and sprites for 2D game environments with \textit{Perlin noise} \cite{Korn2017}.

\subsection{Pattern based}

\textit{Pattern based} approaches use serialised logic to solve generative tasks. Examples include growth or simulation algorithms such as: \textit{cellular automata}, \textit{space colonisation}, \textit{erosion} and \textit{diffusion/propagation}. As well as grammars: \textit{L-systems}, \textit{shape grammars}, \textit{split grammars}, \textit{graph grammars}.
\textit{Cellular automata} has long existed, with early work of Von Neumann \cite{von1966theory} and Conway's Game of Life \cite{gardner1970fantastic}. Cellular automation works on the basis of adjacency rules that determine the value of a cell in a discrete grid. Evaluating each cell in the grid at each step allows for natural growth or formation of shapes and volumes. Cellular automation is an effective approach to content generation, given the appropriate rules. For example, while using a constrained-growth approach to generate floor plans, Green et al. \cite{Green2019} use cellular automata to arrange the placement of windows on walls. Cellular automata is also applied in combination with L-systems for the generation of caves \cite{Antoniuk2016}. Here, cellular automation is used to refine and smooth out the cave formations.
\textit{Space colonisation}, attempts to mimic a natural growth process for branching tree-like shapes. For example, \cite{Ratul2019} use space colonisation for the real-time generation of trees, and \cite{Guo2020} combines multi-view depth data with a rule-based system to perform space-colonisation. Stylised, plant-like designs, based on existing meshes, have also been generated using a form of space colonisation \cite{Zhang2019}, and space colonisation is used in the generation of road networks \cite{DiasFernandes2018}, where it is used as a flexible method for generating organic looking road layouts that conform to user defined constraints. 

Realistic terrain details can be generated using \textit{erosion} simulation. AutoBiomes \cite{Fischer2020} is a pipeline for generating 3D environments with varying biomes. Using a combination of climate simulation, biome refinement and asset placement, an initial noise-based terrain is built upon to create a complex environment. This climate simulation models temperature, wind and precipitation. Franke and Müller \cite{Franke2021} generate cave geometries using simulated physical properties, such as water flow and erosion. This produces a voxel-based volume given a set of parameters. A surface is formed from this volume via marching cubes, and textures are generated using a combination of Perlin and Worley noise.

\textit{Deformation model} is used for generating creatures and trees. In \cite{Dvoroznak2020}, users are able to draw or trace from reference material on a 2D canvas to create creatures. Users build semantic layers of the subject and order them based on depth. These layers are then inflated to form a 3D shape, aided by a deformation model. In \cite{Wang2018c}, graph representations are used for interpolation of tree models from existing examples.

\textit{Deprojection} approaches have been applied in the reconstruction of props and scenes. Fedorov et al. \cite{Fedorov2019} propose single-image 3D mesh generation using edge detection in combination with user demarked guides. This method also generates textures by modifying the input image content.

With the current availability of stereo cameras and the Kinect, RGB-D data can be produced readily at a much lower cost than traditional scanning methods. This added depth information can be instrumental in the task of reconstructing 3D objects, as demonstrated in \cite{Gao2019}. Effective RGB-D scanning pipelines are also suggested by \cite{Slavcheva2016, Niemirepo2021}, performing on par with other state-of-the-art photogrammetry approaches, while \cite{Kim2022} introduce a framework for simultaneously texturing and reconstructing scenes using RGB-D data. Alternatively, \cite{Chen2018} suggest a view-dependant texturing approach to real-time rendering.

In a traditional game development pipeline, 3D art may be produced with \textit{concept art} as reference. It is also common for 2D designs to present different views of the object, typically the front, top and side. In the manual 3D modelling process, these references allow for the accurate reconstruction of designs. This process can be automated by projecting multi-view concept art onto a voxel volume \cite{SilvaJunior2018}, refining and converting the result to a mesh via marching-cubes \cite{Lorensen1987}.

\textit{Propagation} approaches involve the progressive growing of shape through a space. For example, hair generation is primarily achieved by growing strands through a 3D volumetric flow field, representing the directional flow of hair through space \cite{Saito2018, Zhang2018, Zhang2019b, Shen2021a}. Gao, Yao and Jiang \cite{Gao2019} segment individual object meshes from a scene by propagating user assigned labels, capturing the full shape of each object in the scene, while Dijkstra's algorithm is used to traverse a terrain and form natural height variation \cite{Golubev2016}.

Patch-based generation is applied to road networks and web design. In the method of \cite{Teng2017}, main roads are built using a graph growing algorithm. The spaces between the main roads are then  populated with semantically tagged road patches, propagating inwards from the main roads. Instead, Mockdown \cite{Lukes2021} learns layout constraints for positioning web elements.

Deriving from the work of Chomsky \cite{Chomsky1965}, \textit{generative grammars} operate as formalised rules and structures from which instances can be built. There are many variations of grammar employed in the generation of assets, including: \textit{L-systems}, \textit{Shape grammars}, \textit{Graph grammars}, \textit{Split grammars} and \textit{Stochastic grammars}.
Devising a grammar that encompasses the aspects of a reconstruction target, while mapping an input to said grammar is challenging. Li et al. \cite{Li2018a}, propose a probabilistic context-free grammar (PCFG) for reconstructing buildings from images, learning rules from existing models, and \cite{Martinovic2013} introduce an approach to grammar learning which focuses on facades. Alternatively, \cite{Cao2017} attempt to reconstruct facades from low-resolution images, while \cite{Jesus2016} employ a layered approach using grammars.

The challenge of generating buildings with curved surfaces can be addressed with the use of different coordinate systems, allowing for the same grammars to be applied on flat and curved surfaces \cite{Edelsbrunner2016}. Demir, Aliaga and Benes \cite{Demir2016} introduce a method, capable of inferring a grammar from an existing building model. Building on this research, a framework has been established for the proceduralisation of existing 3D models \cite{Demir2017}. Grammars have also been applied in the generation of ancient Roman and Greek style structures \cite{Konecny2016}. Nishida, Bousseau and Aliaga \cite{Nishida2018} reconstruct 3D models of buildings from single-view images, using a series of CNNs that output shape grammar parameters. This is adapted as a web tool \cite{Bhatt2020}, in which users interact with a web-based UI and the bulk of the computation is completed remotely. A sketch-based approach is also explored, allowing users to draw in aspects of a building \cite{Nishida2016}.

Grammars have also been applied in interior layout generation, for example a stochastic grammar with Spatial And-Or Graph (S-AOG) is introduced \cite{Jiang2018}. This method allows for a large degree of user control, while adhering to rules and characteristics that are present in pre-existing data. This can be used to synthesise data for training or validating deep-learning methods. Freiknecht et al. \cite{Freiknecht2020} introduce an algorithm for generating full building assets including interiors and textures. Alternatively, a Scene Grammar Variational Autoencoder (SGVAE) approach is introduced \cite{Purkait2020}, which encodes indoor scene layouts via a grammar.

To create a logo, designers must spend time developing ideas, and creating many variations to find the ideal design. Li, Zhang and Li \cite{Li2017} attempt to alleviate the amount of manual exploration for designers, introducing a framework that augments the logo design process with the use of shape grammars. The Procedural Shape Modeling Language (PSML) \cite{Willis2021} allows users to express 3D shapes via code. This language integrates shape grammars and object-oriented programming to allow object structures to be expressed hierarchically, with adjustable parameters.

Generative approaches may be used for design ideation, where users are not necessarily interested in polished outputs, but rather unique ideas that can be refined. The approach of \cite{Alcaide-Marzal2020} uses a generative grammar system that produces variations of products by combining pre-defined design elements and applying transformations to them. Alternatively, shape grammars have been applied to large scale industrial designs, where parallels are drawn between engineering specifications and generative rules \cite{DosSantos2017}. 
Geometric graph grammars (\textit{GGG}s) are applied to the generation of road networks \cite{Fiser2016}. The \textit{GGG} extends the concept of a graph grammar by encoding geometric data alongside topology.

Volumetric terrain generation can be achieved via voxel grammars \cite{Dey2018}. As a form of shape grammar, voxel grammars define rules that are applied to a given starting point. Alternatively, \cite{Raies2021} combine grammars with a swarm algorithm to generate entire environments consisting of terrain, vegetation and bodies of water. \textit{Swarm grammars} \cite{VonMammen2009}, are devised for each of these aspects, interacting with one another to produce a natural environment.

L-systems are a form of grammar that attempts to model natural growth of plants via recursive re-writing \cite{LINDENMAYER1968280}. L-systems have been adapted to three-dimensions \cite{Hidayat2020} and applied to tree reconstruction from photos \cite{Guo2020}. They have also been applied in road generation \cite{Sharma2016}, and combined with IGA for novel 3D shape generation \cite{KiptiahBintiAriffin2017}.

\subsection{Deep learning}
For the purposes of distinguishing high-level structures from network specifics, generative deep-learning approaches are categorised by their \textit{methods}, and the \textit{architectures} that they employ. In line with the rest of this review, this section will focus on the high-level strategy of each approach. There are two dominant generative deep-learning strategies: GANs, and encoder-decoder networks. These strategies are combined in the form of adversarial auto-encoders (AAEs), and to a lesser extent, deep generative reinforcement learning (RL) and imitation learning (IL) have also been attempted.

Generative deep-learning aims to extract patterns from large datasets, in order to derive novel content. GANs, first introduced by \cite{Goodfellow2014}, are a generative unsupervised learning method that pits two models against each other, such that both models learn through competition. The adversarial strategy has been highly popular in generative approaches throughout the years, with simple image generation \cite{Fadaeddini2018, Liu2019, Karp2021}, sketch-based techniques \cite{Kuriakose2020, zhang2020a} and text-to-image generation \cite{Ramesh2022, Saharia2022, Yue2019, Schulze2021, Peng2021, Qiao2019}.

Basic GAN based implementations have been applied in the generation of spell icons \cite{Karp2021} and textures for games \cite{Fadaeddini2018}, aiding the process of design ideation \cite{Liu2019} and generating images of indoor scenes \cite{Wang2016}. Such generation is \textit{seeded}, as content is produced by randomly sampling the extracted feature space in order to find novel content.

GANs have been successfully applied in style-transfer tasks \cite{Gao2020, Barzilay2021, Shu2021, Yuan2020}, as well as image generation from sketch-input \cite{Kuriakose2020}, art colourisation \cite{zhang2020a}, and caricature generation \cite{Hou2021}. An influential framework for such approaches Pix2Pix \cite{Isola2017}, employs a conditional GAN (cGAN). cGANs take additional inputs, allowing the end user to specify the kind of result generated. For example, \cite{Wang2019} conditions a face image generator on high level attributes, such as age, gender and hair colour. CONGAN \cite{Heim2019} provides an alternative method of input for GANs, in which the user provides photo constraints to the generator, causing it to generate results more like, or less like other images. Instead, StyleGAN \cite{Karras2021} introduces a style based GAN architecture to great effect, producing high quality images of various types.

LayoutGAN \cite{Li2021d} achieves 2D layout generation by learning to produce a feasible layout from a given input. This is further developed with the addition of attribute conditioning \cite{Li2021e}. User-sketch based cGAN approaches have also been applied to the generation of height map based terrains \cite{Panagiotou2020, Wang2020b, Spick2019}. For example, Sketch2Map \cite{Wang2020b} allows designers to draw simple maps that represent terrains, while a similar sketch-based approach, \cite{Du2019}, allows users to sketch sections of terrain that are seamlessly joined. Zhang et al. \cite{Zhang2020} introduce a GAN based method for combined sketch and text based generation, in which the user sketches the shape of the object they wish to depict, then describe its features and colours in text. Existing GAN based attempts at generating novel images using text inputs include \cite{Yue2019} CAGAN \cite{Schulze2021}, SAM-GAN \cite{Peng2021}, CycleGAN \cite{Zhu2017}, MirrorGAN \cite{Qiao2019}, LeicaGAN \cite{Qiao2019a} and ControlGAN \cite{Li2019}. These approaches extract semantic meaning from input-text and apply appropriate transformations to an image via GAN based architectures.

DALL-E \cite{Ramesh2021} and CogView \cite{Ding2021} demonstrate that high quality results can be obtained by scaling the number of parameters and training data to a large degree. Recent methods, such as DALL-E 2 \cite{Ramesh2022} Imagen \cite{Saharia2022}, make use of diffusion and CLIP \cite{radford2021learning} mechanisms to generate high fidelity images from text. Stable-diffusion \cite{Rombach2022} successfully applies a diffusion based approach which can be conditioned on text, images and semantic maps.

The usage of GANs also extends to the creation of 3D assets, where they have been successfully applied in the reconstruction \cite{Knyaz2019, Kniaz2020, Sun2020, Lu2020, Zheng2022}, generation \cite{Li2021a, Li2021c, Shu2019} and interpolation \cite{Shu2019, zeng2022lion, Zheng2022} of new mesh, point-cloud and voxel assets. Furthermore, diffusion models have been applied in 3D shape generation \cite{Hui2022, zeng2022lion}. 

The Sphere as Prior GAN (SP-GAN) \cite{Li2021a} is capable of generating point-clouds in a structure-aware manner, while SG-GAN \cite{Li2021c} and HSGAN \cite{Li2021b} generate point-clouds in topologically and hierarchically aware manners respectively. Voxel generation is also achieved via GANs \cite{Wang2018a, Singh2019, Spick2020}, and cGANs \cite{Ongun2019}. cGAN has also been applied to generating varied voxel-based rock shapes with user defined boundaries \cite{Kuang2019a}.

Hertz et al. \cite{Hertz2020} introduce an approach to shape texture transfer. Given a reference and target mesh, this method is capable of outputting new geometry that applies the texture of the reference to the form of the target mesh, improving on the results of OptCuts \cite{Li2018c}, which instead makes use of 2D displacement maps.

3D model conception can be a combined effort between user and machine. Davis et al. \cite{Davis2021} introduce a VR based co-creative AI which allows users to generate 3D models by exploring and iterating upon ideas. Deep generation of 3D meshes is a difficult task due to limited data availability. This challenge can be avoided with the use of differentiable rendering. Differentiable renderers allow for self-supervision in 2D to 3D tasks, removing the need for 3D ground-truth data. This has successfully been applied to single-view reconstruction \cite{Kato2018, Henderson2020, Pavllo2020} and 2D to 3D style-transfer \cite{Kato2018}, improved upon with use of normal maps \cite{Xiang2019}, and applied in game character face generation from photographs \cite{Shi2022}. With a similar approach, GET3D \cite{gao2022get3d} achieves high quality textured meshes with complex typologies over the full range of 3D asset types. 

In the task of building generation, effective GAN implementations have been presented for internal room layouts using graphs \cite{Nauata2020}, and façade image generation \cite{Sun2022}. These methods are centred around the 2D domain, however, may still be applicable in combination with other methods for generating 3D buildings. Alternatively, \cite{Du2020} present an approach that is capable of generating fully textured 3D building models by chaining multiple GANs together.

Encoder-decoder network structures allow for a mapping, and therefore, translation between input and target domains. Such networks constitute an encoder network, that learns to condense an input, and a decoder network that learns to interpret an output from this embedding \cite{cho2014properties, minaee2021image}. Autoencoders are a form of encoder-decoder that aims to produce outputs that are identical to the input, they are applied primarily in de-noising or compression tasks where the encoder discards irrelevant information \cite{Pascal2010}. U-net is a popular form of encoder-decoder \cite{Ronneberger2015}. As a fully convolutional architecture, it is primarily used for working with images, and as such, U-nets have seen use in diffusion based generative models \cite{Rombach2022, Saharia2022}, image translation \cite{Isola2017}, and interpreting sketches \cite{Delanoy2018a, Kuriakose2020}. 

In general, encoder-decoder networks have been successfully applied in photo-based \cite{Wang2018, Kato2018, Naritomi2021, Lei2020, Pontes2019, Yang2021}, sketch-based \cite{Delanoy2018a, Shen2021, Delanoy2019, Yang2021, Das2021} and text-based 3D shape generation tasks \cite{Liu_2022_CVPR, Canfes_2023_WACV, Mittal_2022_CVPR, Sanghi_2022_CVPR, Jain_2022_CVPR, Liu2022, Fu2022, Yu2022}.

Simple encoder-decoder networks do not learn a consistent latent space that can be sampled from directly to generate new content. Variational autoencoders (VAEs) address this by employing regularisation during the training process, allowing for smooth interpolation and parametrisation of inputs. VAEs have been used to successfully generate 3D assets including furniture \cite{Jones2020, Jones2021, Gao2019a, Gao2021, Cheng2022, Yang2022, Li_Liu_Walder_2022}, textures \cite{Gao2021}, characters \cite{tan2018variational} and hair \cite{Saito2018}.

Jones et al. \cite{Jones2020} introduce a method for generating primitive based objects using a VAE. The model is trained to produce programs in an intermediary language called ShapeAssembly, whereby 3D models are encoded as a list of operations applied to simple cuboids. This is expanded upon with the introduction of a method that is capable of learning macro-operations from existing ShapeAssembly programs \cite{Jones2021}.

SDM-NET \cite{Gao2019a} employs VAEs in learning the structure and geometry of objects, producing high-quality generated and interpolated results. This is developed further with TM-NET \cite{Gao2021}, which produces textured 3D models. A VAE is applied in the generation of novel object and character poses \cite{tan2018variational}, this incorporates the rotation-invariant mesh difference (RIMD) data representation \cite{Gao2016}, allowing the VAE to learn surface deformation.

Deep-learning has been applied in the estimation of hair flow fields. Single-image hair reconstruction is achieved in this way using VAE \cite{Saito2018} or GAN \cite{Zhang2019b}. DeepSketchHair \cite{Shen2021a} instead utilises a sketch-based approach, where users sketch the outline of a hairstyle, and draw lines within to indicate the flow of hair. The user is then capable of refining their design by providing more sketches at different angles, while viewing the result. Alternatively, \cite{Zhang2018} generate hair using multi-view RGB-D images.

Single-view reconstruction allows for the generation of highly specific 3D content with minimal user input. This is largely a task of inferring a whole shape from a single viewing-angle, which can be achieved through prior knowledge of similar objects. Many methods apply deep-learning to the task, such as \cite{Ling2021, Xu2021, Klokov2020, Naritomi2021, Hui_2022_CVPR}.

Pixel2Mesh \cite{Wang2018b, Wang2021}, uses a CNN in combination with a graph convolutional network (GCN) to deform a base mesh with the goal of matching an input image. As this approach manipulates vertices directly, each vertex can also have an associated colour value, which enables the generation of coloured meshes. This architecture has been successfully expanded with a graph attention mechanism \cite{Dongsheng2020} and for multi-view reconstruction \cite{Wen2022}.

There are many other approaches that incorporate the template deformation concept. For instance, Image2Mesh \cite{Pontes2019} encodes an input image using a CNN, finds the closest base model, then deforms it via free-form-deformation \cite{sederberg1986free} to match the input. A similar approach uses free-form-deformation to generate lung models from single-view images \cite{Wang2020}, while template mesh deformation is applied to liver \cite{Tong2020} and heart \cite{Kong2021} reconstruction.

EasyMesh \cite{Sun2020}, also takes a deformation approach, while processing input images into silhouettes to gain consistency in training. Similar approaches use deformation to generate meshes \cite{Li2020, Li2020b, Pan2018}, and point-clouds \cite{Yuniarti2021}. These approaches are limited to deforming existing topologies. Instead, Mesh R-CNN \cite{Gkioxari2019} is capable of reconstructing varying topologies from single-view images. This approach expands om Mask R-CNN \cite{He2017} by adding mesh prediction. The template meshes typically used in such methods are \textit{genus-0}. This covers a wide range of possible geometries, but is not sufficient for all shapes. Pan et al. \cite{Pan2019} circumvent this limitation by introducing topology modification modules which remove faces from the mesh to form holes where they are needed.

One core challenge with single image reconstruction, is the lack of information about the parts of the object that are occluded or facing away from the camera. This issue is addressed in the framework of \cite{Lu2020}, which utilises generated multi-view silhouette data. Yang, Li and Yang \cite{Yang2021} disentangle shape and viewpoint in their encoder-decoder architecture by decoding using separate shape and viewpoint transformer networks.

Generalisation is another challenge, often necessitating copious amounts of data to achieve. One way to circumvent this is to employ a few-shot approach, where a network is capable of generalising to new classes, provided a few samples \cite{Wang2020c}. Lin et al. \cite{Lin2021a} make use of few-shot learning in single-view point-cloud reconstruction by separating class-specific features and class-agnostic features.

Voxel representations have also been used in reconstruction tasks. Due to their structure, they can be interpreted with CNNs, but typically require a large amount of memory to work with at high resolutions. Furthermore, small errors in generation can result in noise. Xie et al. \cite{Xie2018} address this latter issue with a novel weighted voxel representation.

Adaptive O-CNN \cite{Wang2018} is a patch-based approach to 3D object representation using octrees.

Deformation of template meshes is also applied to photo reconstruction for characters \cite{Zhang2021, venkat2019humanmeshnet, Yu2023}, and hands \cite{malik2020handvoxnet}. Additionally, human body meshes have been reconstructed from photographs for the purpose of testing clothing fit in e-commerce \cite{Abdelaziz2021}.
Parameter based face generation is a common technique in games, allowing users to adjust features of a character, this is often achieved using 3D Morphable Models (3DMMs). Deep-learning approaches have succeeded in translating photographs to these parameters \cite{Shi2022, Lin2021b}. Furthermore, the approach of \cite{Lin2021b}, is capable of extracting texture from an input image, and applying it to a reconstructed face model. Fan et al. \cite{Fan2020} introduce a pipeline for full head and face reconstruction based on 3DMM, and, \cite{Ji2020} use a Siamese encoder-decoder architecture for face reconstruction. While 3DMM models represent face shape well, they typically lack fine detail. This is addressed in the method of \cite{Khan2021}, where meshes are refined via displacement, and \cite{Kuang2019b}, where a GAN produces a depth map for a given image. Face generation can be achieved via GAN, such as \cite{Kuang2019b}, or PGAN \cite{Li2019}, which generates faces with the use of geometry images \cite{Gu2002}. The approach of \cite{Shamai2019} employs geometry images and a GAN architecture based on progressive growing GANs \cite{Karras2018}. StyleGAN \cite{Karras2021}, has also been expanded upon for learning 3D aware face generation \cite{Or-El_2022_CVPR}.
In the method of \cite{Li2021f}, a caricature mesh and texture are extracted from a single input photograph and combined. The texture is extracted using a GAN, and facial reconstruction is performed using the method of \cite{Deng2019}. Facial landmarks may be used to improve the accuracy of facial reconstruction methods. Cai et al. \cite{Cai2021b} introduce a method for automatically detecting these landmarks for caricatures.
Some approaches take sketches as input, for example, in the method of \cite{Delanoy2018a}. This method struggles with reconstructing thin structures, due to the resolution of the voxel space. This is addressed by combining voxels with normal maps \cite{Delanoy2019}, where an additional normal prediction network, based on Pix2Pix \cite{Isola2017}, produces normal maps from the input sketches. The normal maps are projected onto the voxel representation, refined, and used to generate a mesh that is far smoother than the previous results. Yang et al. \cite{Yang2021b} introduce a method for generating human body meshes from sketches, and a CNN is applied in converting sketch data to procedural model parameters \cite{Huang2017}.

Some research attempts to extract maximal information from a scene, attempting to either understand a scene holistically or identify and extract individual items. CDMD3DM \cite{Li2021}, for example, reconstructs small scale indoor scenes using RGB-D data as an input, and \cite{Jeon2016} produce accurate texture maps for RGB-D based scene generation. Full scenes have been reconstructed from single-view images, using cGAN architectures \cite{Knyaz2019, Kniaz2020} and CNN based approaches \cite{Weng2021}. Some methods conceive scenes and objects using GAN based approaches, including point-clouds of outdoor scenes \cite{Shinohara2021}, entire cities from single-view images \cite{Kim2020}, and voxel-based scenes that are segmented by object class \cite{Singh2019}.

The placement of furniture within indoor spaces requires an understanding of the functional relationship between furniture types. Approaches to indoor layout generation largely apply deep-learning \cite{Li2019b, Zhang2020c, Jezek2020} and case-based reasoning \cite{Song2017, Song2019} to the task.

AAEs combine the encoder-decoder concept with the the adversarial mechanic of GANs. For example, \cite{Zhao2017} apply AAE to style transfer, in which the latent representations of content and style images are learned within an encoder-decoder, evaluated by a discriminator. AAE have also been applied to interior object placement where an encoder-decoder generator is trained in an adversarial manner against a scene discriminator and image discriminator \cite{Zhang2020c}. The image discriminator takes top-down views of the generated and real scenes, providing an extra visual based assessment.

Reinforcement learning involves learning a policy for a given task. This is achieved by placing agents within an environment and rewarding or punishing their actions in order to guide the policy. Some attempts at asset generation through reinforcement learning have been made, with approaches such as \textit{double deep Q learning }(DDQN) and \textit{deep deterministic policy gradient }(DDPG). For example \cite{Lin2020}, trains agents to reconstruct 3D objects by performing actions similar to human creators, placing primitives and refining geometry with the goal of matching a target model. To set an initial policy, IL is used in the form of \textit{dataset aggregation} (DAgger). This research shows promise, though more work is needed to achieve the generation of detailed models. DDPG is applied to 2D layout generation \cite{HU2021107269}. In this approach, the network attempts to find an optimal layout for a randomised set of elements.

\subsection{Determine Approaches}
To determine an approach for a given task, the pool of existing generative approaches are labelled based on:
\begin{enumerate}
    \item the asset type they seek to generate,
    \item the technique they implement, and
    \item the type of input they take.
\end{enumerate}
Given that $A$, in algorithm \ref{approachAlg}, represents the pool of possible generative approaches, the task of determining valid approaches is described as a process of filtering $A$ in accordance to the three attributes above. Then, the remaining valid approaches are selected based on user preference.

It is possible for a single approach to not be sufficient for some tasks, particularly when the goal is to produce a large-scale system, with many inter-related outputs. For a hypothetical task of generating unique buildings that have interior layouts consisting of procedurally generated furniture, there may not be an existing approach that can achieve this on its own. Yet, when the task is broken down into object placement, building, and furniture generation, for which there are existing approaches, a combined solution can be formulated; by considering where an output of one approach can be the input of another, a pipeline can be formed.

\begin{algorithm} [h]
\caption{Determining Approaches}
\label{approachAlg}
	\begin{algorithmic}
	\scriptsize 
        \Procedure{Determine\_Approaches}{a, t, in, A[], u}
        \Comment{{\scriptsize INPUT: asset type, technique, input type, approaches, user choices}}
            \State $A = [\forall \; app \in A \mid app.input\_type = in \land app.technique = t \land app.asset\_type = a ]$
            \Comment{\scriptsize Filter choices by input type and technique, f stands for filtered}
            \If {$|A| = 1$}
                \State $approach_{s} = A[0]$
            \Else
                \State i=0
                \While{$i < |A| \land A[i] \neq u.approach$}
                        \Comment{\scriptsize Allow user to choose approach}
                        \State $approach_{s} \gets$ $A[i]$
                        \State i = i + 1
                \EndWhile
            \EndIf
            \State{return $approach_{s}$}
            \Comment{Pass selected approach to next step}
	    \EndProcedure
    \end{algorithmic}
\end{algorithm}

\subsection{Choose Datasets and Train}
Due to the prevalence of deep-learning approaches in the literature it is necessary to consider data requirements in regard to training a model appropriately. Supervised and unsupervised deep-learning approaches typically require large labeled or unlabeled datasets, of which many exist publicly, such as ShapeNet \cite{Chang2015}, PartNet \cite{Mo2019} or MPI FAUST \cite{Bogo2014a}.

Alternatively, reinforcement learning requires hand-crafted training environments and reward functions. Though usage of reinforcement learning as an approach to asset generation is largely unexplored, Lin et al. \cite{Lin2020} propose an RL network that successfully learns 3D modelling policies. This approach allows the RL agent to edit a 3D model using typical 3D modelling actions, rewarding it based on the similarity of the result to a target shape. With the usage of target shapes, this approach requires pre-existing data, much like supervised and unsupervised approaches.

\begin{figure}[h]
  \centering
  \includegraphics[width=0.8\linewidth]{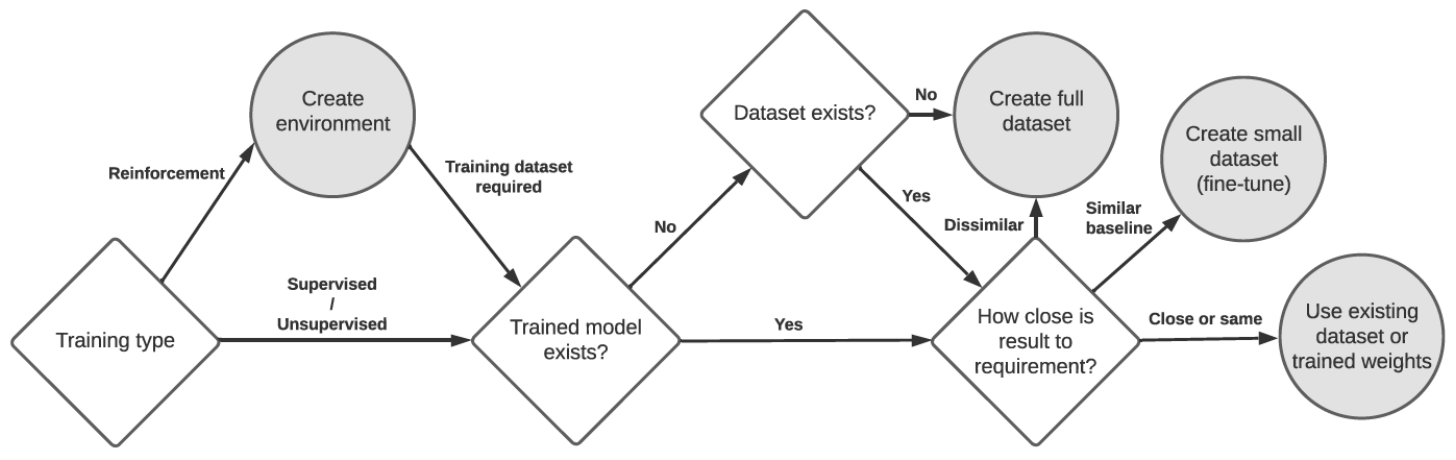}
  \caption{Approach to choosing or creating a dataset.}
  \label{choosingDataset}
\end{figure}

Figure \ref{choosingDataset} presents the proposed process for choosing or creating a dataset in order to train a chosen deep-learning approach. If a supervised or unsupervised learning approach is chosen and the approach is already validated on the target asset type, then the original authors may have provided existing trained weights that can be used. Alternatively, if this is not the case, the original authors may have provided the dataset used to validate their approach. The applicability of existing weights or datasets can be determined by observing their outputs and comparing them to the target result. If they suffice then they may be used. However, if they are close to matching the desired result, a small dataset may be collected and used to fine-tune the model from the existing weights. If no existing weights or datasets are provided, then it may be necessary to obtain an alternative dataset that matches the data requirements of the chosen approach.

\section{Generate assets process} \label{section-generating}
At this stage, the method for generating graphical assets is formulated and ready for implementation. When multiple sub-components are required, multiple approaches are needed. These approaches must be formed into a generative pipeline, with consideration for the necessary order of operation. This is achieved by chaining the approaches, mapping output to input. The order of approaches can be determined by considering each generative approach's pre-requisite data. For example, if the task was to generate a building with an interior containing furniture, the system may begin with a building generator and furniture generator, which both feed into an interior layout generator, as interior object placement requires both a defined environment and objects to place.

Once implemented the generative method should produce assets of the type defined by the user, given the required inputs. Depending on the approaches used, generated assets will be in 2, or 3 dimensions depending on asset type chosen, however there are multiple formats for presenting 2D and 3D data. The user's required format may differ from the format of the output. This can fortunately be rectified using algorithms that convert from one format to another. In the next section the final step in the framework, format conversion, will be discussed.

\section{Formats} \label{section-converting}
\emph{Format} is an important aspect of a graphical asset, determining how the asset may be used, manipulated and presented on-screen. Each possible format has limitations and benefits. In the framework, figure \ref{frameworkFull}, graphical asset formats are presented under two categories: \emph{2D} and \emph{3D}.

3D data can be presented as volumes, or a surfaces. Voxel representations are volumetric. This means that they determine the space that an object occupies. They are the primary format for presenting 3D shapes by volume, allowing for overhangs and tunnels in terrains \cite{Dey2018, Sin2018}. It is also common to convert voxel data to mesh data using marching cubes \cite{Lorensen1987} or surface nets \cite{Gibson1998}. Surface representations such as meshes or point-clouds instead directly represent the outline of an object. Meshes are a common format for 3D content in games and other visualisations. A watertight mesh is a mesh that has a complete, connected surface, such as in \cite{Kazhdan2006, Davis2021}. This is often a requirement for 3D content in real-time rendering and 3D printing. Alternatively a polygon soup lacks full connectivity, seen in \cite{Guan2020}. Point-clouds, much like meshes, represent surfaces of objects or environments, and are used in the perception of real-world space, being applied to computer vision among other tasks \cite{Ongun2021}. Such data is obtained in abundance, due to being the natural output of 3D scanning technology. While point-clouds provide excellent spatial information, they lack structure, and are not typically used directly within digital media applications. Nonetheless, there is a great deal of research into generating point-cloud data \cite{Shu2019, Ongun2021, Yuniarti2021, Lim2022, Li2021c}.

2D assets take the form of either bitmap or vector graphics. Depending on the application and art-style, one form may be chosen over the other. Vector graphics benefit from being procedural, thus constituting comparatively smaller file sizes, and limitless levels of detail, though they are constrained to more simple or block-colour art-styles as a result. Bitmaps, however, allow for more expression in terms of art-style, and benefit in particular from CNN based generative approaches. Though some approaches make use of implicit shape representations \cite{Hui_2022_CVPR, Hui2022}, or signed distance fields (SDF) \cite{Zheng2022, Or-El_2022_CVPR, Slavcheva2016, gao2022get3d}, these forms are not used to represent final \textit{useable} assets, and are therefore not deemed graphical asset formats.

\subsection{Conversion Methods}
The output format of a particular generator may not match the user's desired format. In such cases, output data can be converted to another format using a one-to-one conversion method. These methods, unlike the generative approaches previously discussed, aim to translate data from one format to another with minimal loss of information.

The framework, figure \ref{frameworkFull}, presents the conversion methods. These are drawn from conversion methods observed in the literature, though it is acknowledged that other methods may exist. Some methods are named, while others, such as voxelisation and rasterisation refer to general approaches that may vary in implementation, depending on the use case.

Marching cubes is a popular method for converting voxel data into a complete surface mesh \cite{Lorensen1987}. The inverse conversion, mesh to voxel, is achieved through voxelisation \cite{274942}. Similarly, point-cloud voxelisation is achievable using the method of \cite{Hinks2013}. Conversion from mesh to point-cloud can be achieved through random point sampling \cite{meshsamplepoints}, and conversion of point-cloud data to mesh data can be achieved using Poisson surface reconstruction \cite{Kazhdan2006}.

For 2D formats, the process of converting bitmap data to vector data, vectorisation \cite{Dori1999, Song2002}, and the inverse, rasterisation \cite{Scratchapixel2.02022}, can be applied. Conversion from a 3D format to a 2D format can also be achieved by rasterising the asset at a single viewing angle, though the reverse of this cannot be achieved without a photo or sketch-based generative technique, as additional data must be inferred. Instead, conversion from 2D to 3D can be performed through visual hull \cite{Laurentini1994}, which requires multiple images at different viewing angles. Alternatively, deep-learning generative approaches such as \cite{Shen2021} achieve format conversion, though results are less reliable.

Conversion methods may also be used when multiple approaches are employed within a generator, for example one approach may produce a voxel output but the following approach may require a mesh as input. A conversion method may be used in this case to convert the voxel output to a mesh before it is passed to the second approach.

\subsection{Verify/Convert Format}
\begin{figure}[ht]
    \begin{minipage}{0.9\linewidth}
      \centering
      \includegraphics[width=\linewidth]{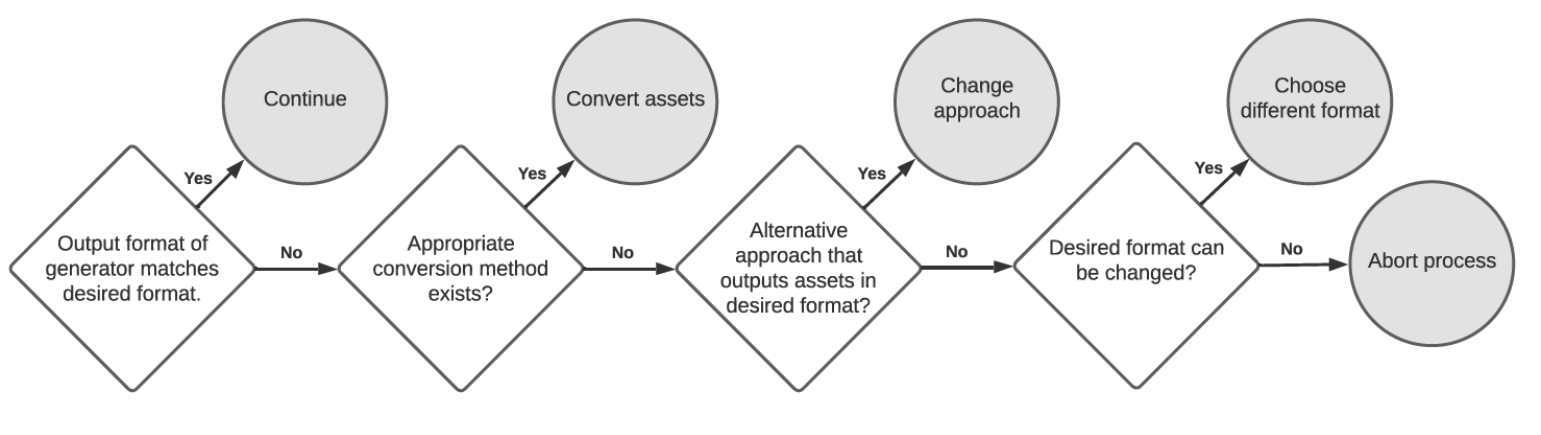}
      \caption{Process for selecting a conversion method.}
      \label{conversionProcess}
  \end{minipage}
\end{figure}

Figure \ref{conversionProcess} presents the process for selecting a conversion method. First the user's desired graphical asset format should be determined. If a conversion method exists between the output format of the approach and the desired format, this may then be used. If the output format of the approach matches the desired format, then a conversion method is not necessary. In the scenario where no conversion method is applicable, the user may consider selecting an alternative approach, or reconsidering the desired format.

\begin{figure}[ht]
  \centering
  \includegraphics[width=0.8\linewidth]{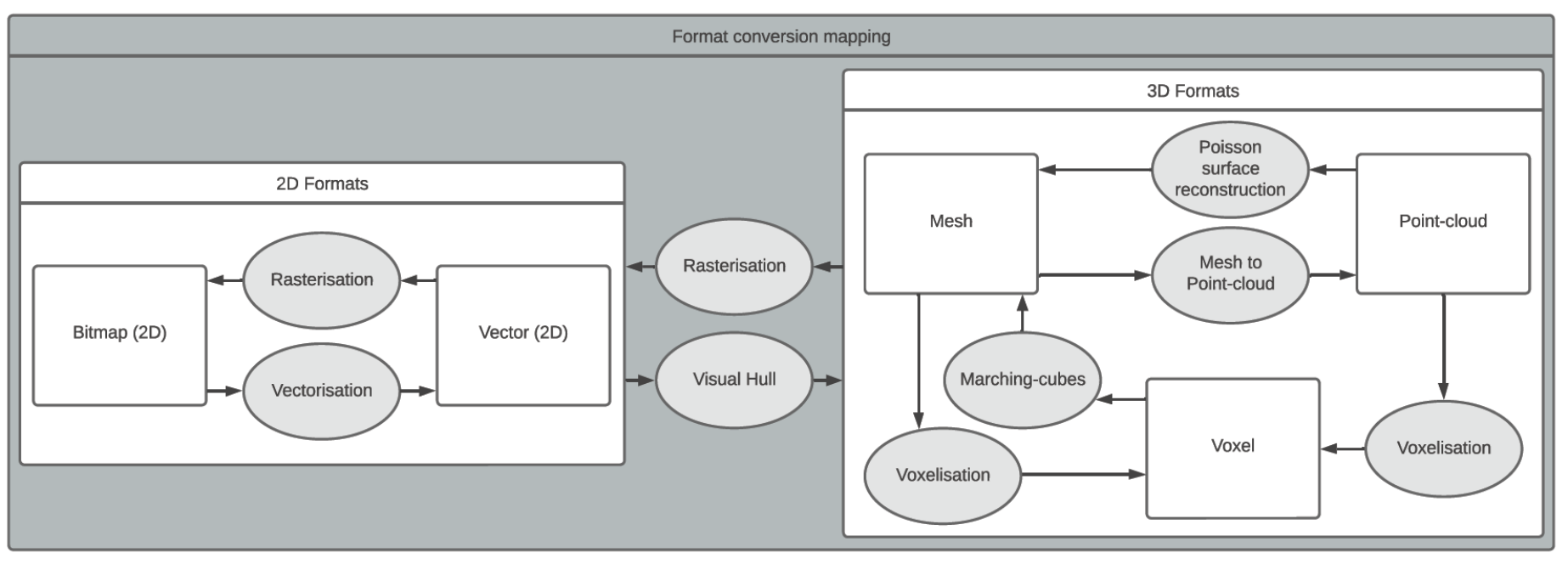}
  \caption{The mapping between conversion methods and formats.}
  \label{conversionMapping}
\end{figure}

Figure \ref{conversionMapping} presents the mapping between formats and conversion methods. Within each set of formats, 2D and 3D, data can be freely converted using a single method. When converting from 3D to 2D, however, it is suggested that the asset be converted to mesh format so that it may be rasterised. Likewise, when converting from 2D to 3D using visual hull \cite{Laurentini1994} any vector graphics must be converted to bitmap.

\section{Concluding discussion}

Current methods for generating and transforming graphical assets are dispersed across a wide array of applications. While this suggests that graphical asset generation is valuable in many contexts, there is no centralised way of finding the available approaches and techniques, and furthermore no standardised process for selecting or devising a method for a particular task.

To address this gap, a systematic literature review was conducted, bringing together the state-of-the-art in generative methods for graphical asset creation. We reviewed 200 papers, in which methods for generating 21 types of asset were explored, identifying 9 forms of technique and 28 types of approach. The GAGeTx framework assembles the findings with a process for building graphical asset generators or transformers based on the needs of the user or practitioner. The goal is to make the breadth of current methods accessible to interested parties, encourage cross-over of techniques and approaches, and ensure the visibility of state-of-the-art capabilities across disciplines.

Though generative methods are rapidly evolving in the direction of deep-learning, there is still much potential in applying existing methods cross-discipline. Areas where large quantities of graphical assets are required, such as game development, may benefit from such generative tools and automation. In this case efforts must be made to establish the needs for such specialised users.

As for GAGeTx more work is required to validate and expand the framework. In its current iteration GAGeTx does not provide a way for selecting the best approach for a task, rather, it leaves this decision up to the user. To address this, future work will include the formation of a method for comparing each generative approach. Furthermore, the framework should be validated through usage. The breadth of techniques and approaches will undoubtedly continue to grow and progress with future advancements, and with it the framework would evolve. Finally, implementation of the framework as a tool may maximise the impact and effectiveness of the framework.

\bibliographystyle{unsrtnat}

\end{document}